# Dual atom ($^{87}$Rb-$^{133}$Cs) grating magneto-optical trap

LEI XU,[1,2] MUMING LI,[2] ZHILONG YU,[2] ZHEYU LIU,[2] JUNYI DUAN,[3,†] FANG WANG,[1,2] FENG ZHAO,[1,2] AND XIAOCHI LIU[1,2,*]

[1]*University of Chinese Academy of Sciences, Beijing, 100049, China*
[2]*Key Laboratory of Atomic Frequency Standards, Innovation Academy for Precision Measurement Science and Technology, Chinese Academy of Sciences, Wuhan, 430071, China*
[3]*Center for Advanced Measurement Science, National Institute of Metrology, Beijing, 100029, China*
[†]*duanjy@nim.ac.cn*
[*]*liuxc@apm.ac.cn*

**Abstract:** This paper proposes a dual-color grating chip design method for simultaneously capturing dual atomic clouds ($^{87}$Rb and $^{133}$Cs). By simulating key parameters such as the grating period, etching depth, duty cycle, coating material, and thickness, the optimal design parameters were determined to ensure efficient dual-wavelength diffraction and maximize the number of captured atoms. Experimental results demonstrate the simultaneous trapping of $1.6 \times 10^8$ $^{87}$Rb atoms and $7.8 \times 10^6$ $^{133}$Cs atoms, thereby offering an approach for multi-species cold atom systems. This dual-species grating magneto-optical trap (GMOT) system has potential applications in precision measurements such as cold atom clocks, quantum interferometers, and quantum electrometry.

## 1. Introduction

Conventional magneto-optical traps (MOT) for neutral atoms typically employ a configuration of six counter-propagating laser beams and anti-Helmholtz coils, resulting in bulky and complex structures. However, the demand for miniaturization and low power consumption in quantum precision measurement systems, such as atomic clocks[1], magnetometers[2], and interferometers[3], has led to the development of compact MOT systems. Several approaches were proposed such as pyramid-shaped MOTs[4], grating MOTs[5][6][7][8], and Fresnel MOTs[9]. These systems utilize a single laser beam incident perpendicular to the device surface, with diffracted light used to cool and trap atoms, thereby considerably simplifying the setup. The grating used for laser cooling is often composed of three linear one-dimensional gratings arranged at 120° to each other[1]. Grating MOTs (GMOT) have successfully trapped on the order of $10^8$ cold atoms, with temperatures reaching the microkelvin range, and their planar characteristics facilitate atom interrogation[10][11][12].

Currently, GMOTs are primarily used to prepare and study single-species atomic clouds. Dual-species systems and dual-color MOTs, however, continue to hold significant influence in the field of quantum precision measurements[13][14][15][16][17][18]. For example, the Rydberg energy levels range of dual-species systems can be expanded and systematic uncertainties can be eliminated[19]. Bondza S. et al.[13] constructed a red-blue dual-color MOT on a single grating to achieve two-stage cooling of $^{88}$Sr. Schioppo M. et al. [14] used two atomic systems to suppress the Dick effect caused by the dead time.

Therefore, we tend to use grating to develop dual atomic clouds simultaneously including rubidium and cesium, which have similar vapor pressures at room temperature, making simultaneous operation under similar vacuum conditions feasible. Both Cs and Rb belong to the alkali metal group, sharing similar properties, and are representative in cold alkali metal atom research. For instance, $^{133}$Cs and $^{87}$Rb are used as the primary and secondary definitions of the second in the international time standard[20][21]. In the field of quantum sensing, Rb and Cs are also commonly used in electric field measurements[22][23]. However, we cannot simply apply single-species gratings to develop dual-color MOTs or dual-species (such as Rb, Cs) atomic clouds, as the second species cannot be trapped because of differences in the diffraction efficiency and diffraction angles for various cooling wavelengths on the same grating, thus resulting in non-ideal conditions for both species. This paper proposes a design method for a dual-color grating chip for capturing dual-species atomic clouds. The structure of the grating is defined to determine a set of parameters that correspond to the appropriate diffraction rate with the goal of maximizing the number of atoms theoretically calculated. Experimentally, we fabricated this designed grating chip and used it to trap $^{87}$Rb and $^{133}$Cs simultaneously, thereby offering guidance for future experiments involving multispecies atom cooling and trapping in GMOT systems.

## 2. Computation and simulation of grating for MOT

The diffraction efficiency of the grating chip is the most dominant parameter for simulations[13][24][25][26][27][28]. This section defines the grating chip in terms of its period *d*, etching depth *T*, duty cycle *r* (the percentage of the grating period that is not etched), coating material, and thickness *h*, as shown in Figure 1(*s* is the effective length of the etched part), to compute and simulate the efficiency. Before the simulation, the scanning parameter ranges must be determined based on several grating diffraction computing formulas. The grating equation and the theoretical diffraction efficiency are defined as follows:[26]

$$d \sin \theta = m\lambda \quad (1)$$

$$\frac{P_0}{P_{in}} = \rho[1 + 2r(r-1)(1 - \cos(\frac{4\pi T}{\lambda}))] \quad (2)$$

$$\frac{P_m}{P_{in}} = \frac{\rho}{m^2\pi^2}[\sin^2(m\pi r) + \sin^2(m\pi s/d) + 2\cos(\varphi)\sin(m\pi r)\sin(m\pi s/d)] \quad (3)$$

Equation (1) is the grating equation for the vertical incidence, equation (2) defines the zero-order diffraction efficiency, and equation (3) defines the m-order diffraction efficiency. Equations (2) and (3) show that when *r* = 0.5, the zero-order diffraction efficiency reaches a minimum, and the first-order diffraction efficiency reaches a maximum. Equation (1) shows that when λ < *d* < 2λ, no second-order diffraction exists, and the first-order diffraction is retained. Therefore, we selected a grating period that satisfies both wavelengths, that is, 852 nm < *d* < 1560 nm. Based on the work by Nishii et al.[5], an etching depth of *T* ≈ λ / 4 ensures a large first-order diffraction efficiency; therefore *T* = 175 – 215 nm.

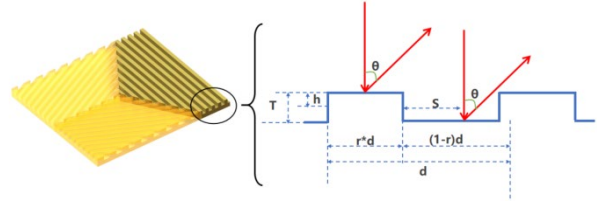

Fig. 1. Schematic diagram of grating structure(2D)

### 2.1 Simulation results and analysis

Based on the above calculation, *r* is set to 0.5, and *d* and *T* have been preliminarily determined. However, equations (1), (2), and (3) are general expressions for the diffraction efficiency of uncoated gratings. As our grating requires wavelength-dependent high-reflective coating, the effect of the coating parameters on the diffraction efficiency must be considered, which then requires a re-sweep of parameters near the initial values.

Gold (Au) is commonly used as a high-reflectivity material for reflective gratings owing to its resistance to oxidation[10][29]. Its reflectivity varies with wavelength; for near-infrared wavelengths, its value is generally above 0.9, whereas it is slightly lower at 780 nm than 852 nm. Considering the fabrication difficulty and cost, we set the coating thickness *h* to 100 nm.

Further sweeping the etching depth *T* and grating period *d* yields the diffraction efficiency variation for 780 nm and 852 nm, as shown in Figure 2. According to the dual-color results, within the ranges of *T* = 175 – 215 nm and *d* = 950 – 1450 nm, the diffraction rate remains relatively stable and falls within the range of 0.38 – 0.44.

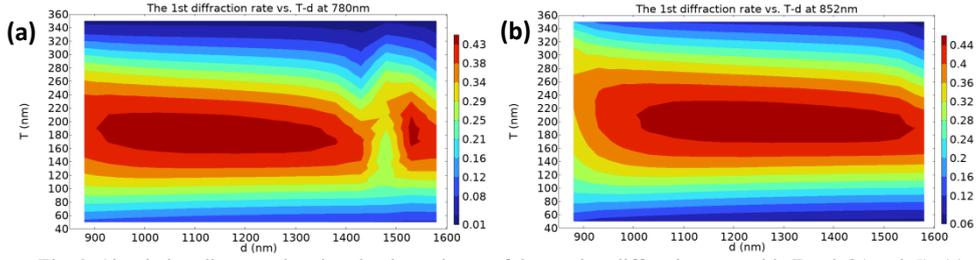

Fig. 2. Simulation diagram showing the dependence of the grating diffraction rate with $T$ and $d$ ($r = 0.5$). (a) Variation in the diffraction rate with $T$-$d$ at an incidence of 780 nm and (b) 852 nm.

Burrow et al.[27] proposed the concept of a balance factor, which is expressed by equation (4), where $\eta_1$ is the first-order diffraction efficiency, $\eta_0$ is the zero-order diffraction efficiency, and $\theta$ is the diffraction angle:

$$\eta_B = \frac{3\eta_1 \cos\theta}{1-\eta_0}. \tag{4}$$

The equation suggests that the theoretical balance factor is 1 and the optimal axial force is achieved when the first-order diffraction efficiency is 33%. Substituting the simulated diffraction efficiency results and their corresponding grating parameters from Figure 2 into this equation, we determine that no single set of grating parameters yields a balance factor of 1 and strictly achieves a 33% diffraction efficiency for both 780 and 852 nm simultaneously. This discrepancy is primarily owing to the slightly higher reflectivity of Au at 852 nm, which causes the diffraction efficiency at 852 nm to be consistently higher than that at 780 nm. Moreover, fabrication errors of ±10 nm are common, and as shown in Figure 2, the range of $T$ values that correspond to a diffraction efficiency of approximately 33% is quite narrow. Fabrication errors can lead to significant changes in the local diffraction efficiency. Therefore, based on the simulation results and practical experimental considerations, we expanded the parameter selection range slightly, choosing balance factors between 0.95 and 1.05, and diffraction efficiencies between 36% and 44% as the candidate parameters. Figure 3 shows the balance factor as a function of the grating parameters and their corresponding diffraction efficiencies. The red-marked regions represent the parameters that met our selection criteria; the other regions were discarded. Hence, we selected $d = 1150 – 1300$ nm and $T = 210$ nm (easier to fabricate) as candidate parameters for fabrication.

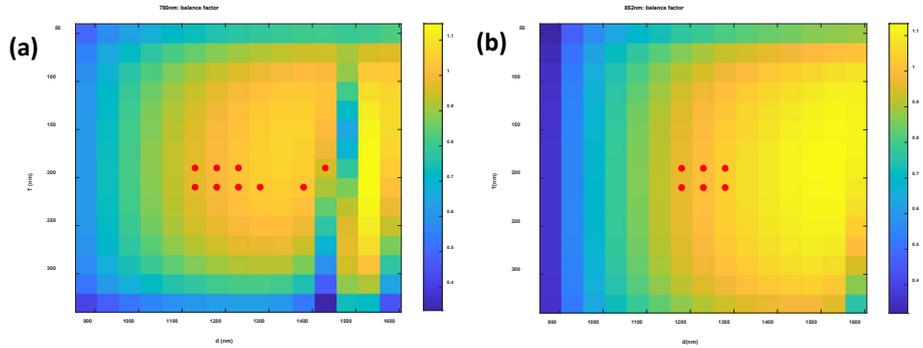

Fig. 3. Diffraction rates obtained from various grating parameters and their corresponding balance factors. (a) Balance factor at 780 nm incidence; (b) Balance factor at 852 nm incidence.

### 2.2 Atoms number estimation

The number of atoms is a crucial performance parameter in cold-atom measurements because a larger number of trapped atoms significantly enhances the signal-to-noise ratio (SNR) in precision measurements. Therefore, the number of atoms must be calculated to determine the grating parameters. The steady-state number of atoms in a trap is given by equation (5)[28]:

$$N = \frac{S}{8\sigma} \cdot \left(\frac{v_c}{v_T}\right)^4. \tag{5}$$

Here, $v_c$ represents the maximum capture velocity of the atoms, below which all the atoms are trapped; $\sigma$ denotes the collisional cross-section between trapped atoms and background gas atoms at room temperature. This equation neglects collisions between cold atoms within the trap because the velocity of laser-cooled atoms is negligible compared to that of background gas atoms, meaning that the loss rate due to cold-atom collisions can be disregarded. For Rb-Rb collisions, the cross-section is approximately $2\times10^{-13}$ cm², and the cross-section for Cs-Cs collisions is comparable to that of Rb[30]. The term $v_T = \sqrt{2k_B T/m}$ represents the average velocity of the background gas atoms and $S$ is the surface area of the trap. The region of the GMOT can be approximated as a combination of two triangular pyramids as illustrated in Figure 4.

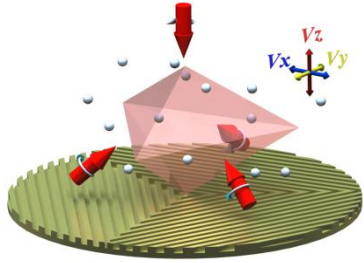

Fig. 4. Shape of the grating magneto-optical trap region

Therefore, $S = 6\sqrt{3} \cdot (\frac{h}{2})^2 \cdot \frac{\tan\theta}{\cos\theta}$, where $h$ denotes the height of the trap and $\theta$ is the diffraction angle.

The components of $v_c$ along the axial and radial directions are defined as $v_z$, $v_x$ and $v_y$, as shown in Figure 4. The forces on Rb and Cs need to be analyzed, following which Newton's second law is applied to compute these components. The transition components in the GMOT are defined by equations (6)–(8)[31]:

$\sigma^+$ transition: $\quad I_{j,+1} = I_j(\frac{1-S_3 \cos\theta}{2})^2$ (6)

$\sigma^-$ transition: $\quad I_{j,-1} = I_j(\frac{1+S_3 \cos\theta}{2})^2$ (7)

$\pi$ transition: $\quad I_{j,0} = I_j \frac{\sin^2\theta}{2},$ (8)

where $S_3$ is the circular-polarization component. Index $j$ corresponds to the number of incident and diffracted beams. Based on these equations, the net force acting on the atom can be expressed by equation (9):

$$F = \sum_{j=1}^{N} \frac{\hbar k_j \Gamma}{2} \sum_{q=-1,0,1} \frac{\frac{I_{j,q}}{I_S}}{1+\frac{I_{j,q}}{I_S}+\frac{4(\Delta-\delta_q)^2}{\Gamma^2}} k_j$$

(9)

where $k_j$ is the wave vector, $\Gamma$ is the natural linewidth, $I_j$ represents the intensity of the incident and diffracted light, $I_S$ is the saturation intensity, $\Delta$ is the laser detuning, and $\delta_q$ is the Zeeman shift.

Based on the results from Section 2.1, we selected the appropriate parameter range ($T$ = 210 nm, $r$ = 0.5, $h$ = 100 nm, $d$ = 1150 – 1300 nm) and their corresponding diffraction efficiencies and set the incident light intensity to one saturation intensity, the detuning to 1.5 times the natural linewidth, and the magnetic field gradient to 8 G/cm, with the circular polarization set to 1. The relationship between the maximum capture velocity and the grating period $d$ is shown in Figure 5.

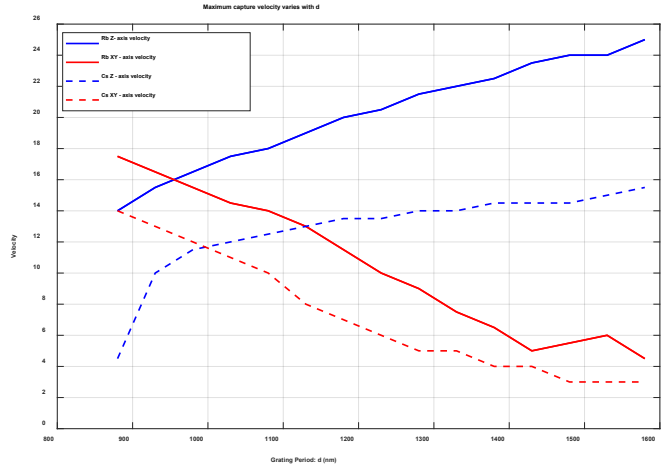

Fig. 5. Variations in the maximum capture velocity with the grating period *d*.

The results in Figure 5 indicate that for both Rb and Cs, the axial capture velocity increases with *d*, while the radial capture velocity decreases with *d*. This trend is observed because a smaller diffraction angle enlarges the axial boundary of the MOT, while reducing the radial boundary, thus making it easier for atoms to escape in the radial direction. In addition, the maximum capture velocities for Cs in all directions are lower than those for Rb. This difference is attributed to three factors: (1) Cs atoms have a greater mass; (2) the natural linewidth of Cs is slightly smaller than that of Rb, meaning that Cs has a longer lifetime in its excited state; (3) the higher reflectivity of the gold coating at 852 nm leads to a greater deviation in the diffraction efficiency from the ideal 33%, which affects the balance factor of Cs. When the capture velocities $v_x$, $v_y$, and $v_z$ are large enough and nearly equal, the atom capture velocity is considered optimal. For example, at $d \approx 1050$ nm, both Rb and Cs achieve their maximum capture velocities.

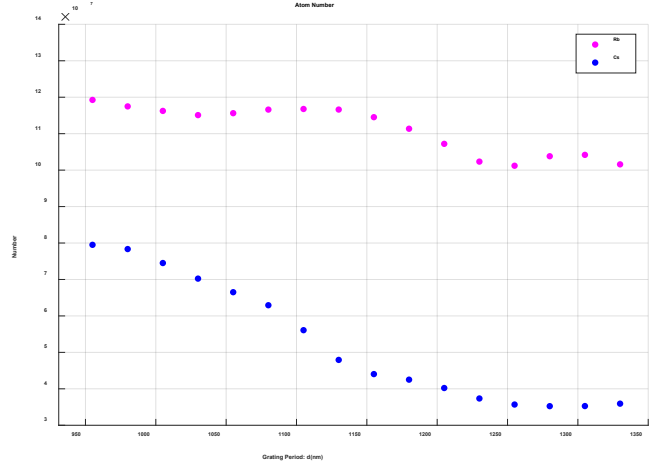

Fig. 6. Estimation of theoretical atomic number as a function of grating period *d*.

Based on the above results, we estimated and found that the number of atoms in the dual-wavelength GMOT for both $^{87}$Rb and $^{133}$Cs can reach $10^7$ when $d = 1150 – 1230$ nm, as shown in Figure 6. Because the maximum capture velocity of $^{133}$Cs is lower than that of $^{87}$Rb, the final atom count for Cs is also lower. To maximize the number of trapped atoms and reduce fabrication difficulty, we selected $d = 1150$ nm.

Eventually, to balance the fabrication complexity and simulation results, we selected $T = 210$ nm, $r = 0.5$, $h = 100$ nm, and $d = 1150$ nm as the final design parameters for the dual-wavelength grating used in the experimental construction of the $^{87}$Rb-$^{133}$Cs dual-species GMOT.

## 3. Experimental results

*3.1 Grating and its optical performance testing*

Based on our design, the dual-color grating chip was fabricated, with a total size of 20 × 20 mm, along with its scanning electron microscope (SEM) images, as shown in Figure 7. The chip was fabricated on a monocrystalline silicon substrate coated with a 100 nm gold thin film. The grating period was 1150 nm with a grid line spacing of 575.1 nm. The optical properties of the grating were characterized using 780 nm and 852 nm lasers, both with a beam diameter of 2 mm and an incident power of 3 mW. A quarter-wave plate and a half-wave plate were used to adjust the incident beam polarization, achieving pure left-handed circular polarization ($S_3 = -1$). Figure 8(a) compares the measured and simulated polarization of the first-order diffracted light on each one-dimensional sub-grating, quantified by the Stokes parameter $S_3$. The measured diffraction efficiency is shown in Figure 8(b) and the diffraction angles are shown in Figure 8(c). Across Figure 8, the black solid lines represent the theoretical $S_3$, diffraction efficiency, and diffraction angle for the 780 nm beam, whereas the black dashed lines represent the 852 nm beam. The red, green, and blue squares correspond to data from different grating regions. Solid squares indicate measurements for the 780 nm beam, while hollow squares indicate the same for the 852 nm beam (five trials).

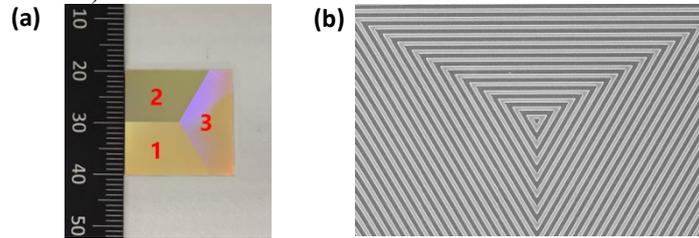

Fig. 7. (a) Overview of grating chip; (b) SEM image of grating chip.

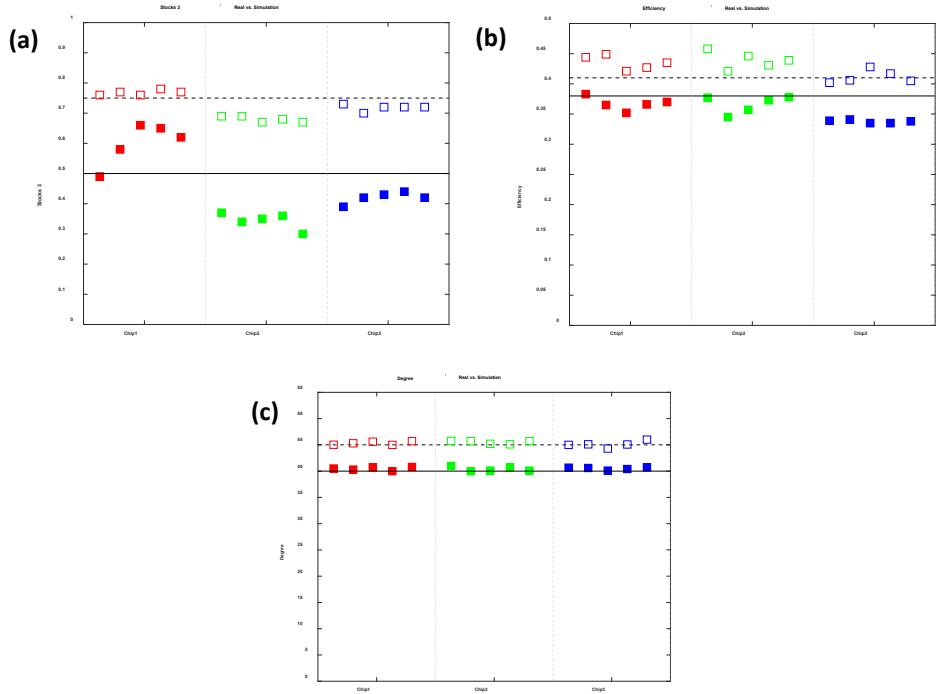

Fig.8. Comparison between simulation (lines) and experimental results (squares). (a) $S_3$; (b) Efficiency; (c) Degree.

The common deviations between the measured grating performance parameters and the simulated values are listed in Tables 1 and 2. Although the measured diffraction angles did not perfectly match the simulated values, the deviation remained within a 10% tolerance range. For the diffraction efficiency $\eta_1$ and Stokes parameter $S_3$ of the first-order beams, discrepancies occurred between the experimental results and the simulations.

Despite these deviations, which can be attributed to fabrication imperfections, the experimental data showed a convergence trend towards the overall simulation.

Table 1. Percentage tolerance of grating performance parameters ε（780 nm）

|  | $S_3$ | $\eta_1$ | $\theta$ |
|---|---|---|---|
| Chip1 | 40.0% | 6.7% | 4.0% |
| Chip2 | 62.4% | 7.4% | 4.9% |
| Chip3 | 32.0% | 22.3% | 3.5% |

Table 2. Percentage tolerance of grating performance parameters ε（852 nm）

|  | $S_3$ | $\eta_1$ | $\theta$ |
|---|---|---|---|
| Chip1 | 4.8% | 12.3% | 3.1% |
| Chip2 | 18.6% | 14.1 % | 3.3% |
| Chip3 | 8.5% | 1.6 % | 4.4% |

*3.2 Simultaneous preparation of Rb and Cs atom clouds using grating magneto-optical trap*

A dual atom GMOT was developed based on our grating chip. We utilized an ultrahigh vacuum (UHV) chamber with a 5 L/s ion pump in which $^{87}$Rb vapor was supplied via a dispenser and $^{133}$Cs was housed in a crushed copper tube and released through a valve. The grating was mounted on the outer wall at the bottom of the UHV chamber. A schematic of the setup is shown in Figure 9. The collimator *A* delivered a Gaussian beam with a waist of 7.8 mm, containing 80 mW of 780 nm cooling light; the collimator *B* delivered a Gaussian beam with a waist of 7.8 mm, containing 82 mW of 852 nm cooling light. Both beams were circularly polarized using quarter-wave plates, combined via a dichroic mirror, and then expanded through a 30 mm lens. The beam area reaching the grating was regarded as a quasi-flat top.

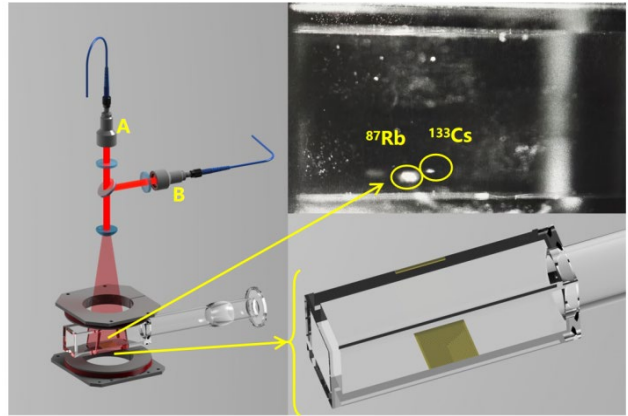

Fig. 9. Schematic diagram of grating magneto-optical trap constructed by vacuum chamber and cooling light.

The applied magnetic field gradient was 8 G/cm. The 780 nm cooling beam was red-detuned by 8 MHz from the $^{87}$Rb D2 transition |F=2> to |F'=3>;. the 852 nm cooling light was red-detuned by 8 MHz from the $^{133}$Cs D2 transition |F=4> to |F'=3co5>. The vacuum degree is maintained around $1.1\times10^{-6}$ Pa. Under these experimental conditions, we successfully captured $1.6 \times 10^8$ $^{87}$Rb atoms and $7.8 \times 10^6$ $^{133}$Cs atoms. The dual-cold atom clouds are shown in Figure 9. Figure 10 shows the relationship between the atom number and the magnetic field gradient, optical power density, and optical detuning. The atoms number reaches the maximum at the magnetic field gradient of 8 G/cm without further increment. This is probably because the effects of light and magnetic fields on different atoms result in different forces and positions. However, the MOT region contains only one zero-point; therefore, the gradient zeros and light-scattering force equilibrium points of the two atomic clouds cannot completely coincide with each other, which results in the number of atoms not reaching saturation. At an optical power density of 15 – 20 mW/cm$^2$, the number of atoms gradually saturates. The number of Rb atoms reached its maximum when the optical detuning was approximately 9 MHz, and the number of Cs atoms reached its maximum when the optical detuning was approximately 8 MHz. We also found that the number of Rb atoms reached the theoretical estimation level in Section 2.2, whereas the number of Cs did not. Two possible explanations can be provided: (1) grating fabrication errors, (2) zero-point offset of light and magnetic fields.

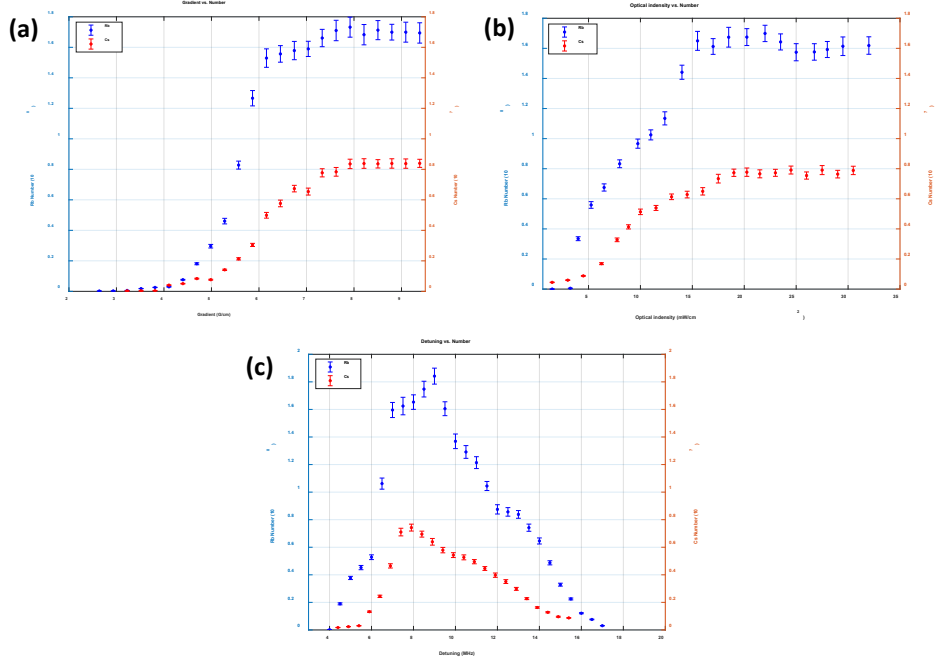

Fig. 10. (a) Atom number versus the magnetic field gradient; (b) Atom number versus the cooling laser optical intensity; (c) Atom number versus the cooling laser detuning.

## 4. Conclusion

We proposed a design method for a dual-color grating chip to capture dual-species atomic clouds. The diffraction efficiency of the grating chip is the most critical parameter in simulations. We defined the grating chip in terms of its period $d$, etching depth $T$, duty cycle $r$, coating material, and thickness $h$, to simulate efficiency. According to the simulation, to guarantee the appropriate efficiency and balance factor, and to maximize the atom number, we determined the following values for the parameters: $T = 210$ nm, $r = 0.5$, $h = 100$ nm, and $d = 1150$ nm – 1300 nm. To balance the fabrication complexity and the simulation results, we finally selected $T = 210$ nm, $r = 0.5$, $h = 100$ nm, and $d = 1150$ nm as the final design parameters for the dual-wavelength grating.

Experimentally, we successfully trapped $1.6 \times 10^8$ $^{87}$Rb atoms and $7.8 \times 10^6$ $^{133}$Cs atoms simultaneously. The experimental results show that the positions of the Cs cloud and the Rb cloud are not exactly overlapped radically. This is partially due to the fabrication uniformity of the different grating zones shown in Figure 7 and Figure 8. Generally, in the single atom specie GMOT, the exact position of the atoms is not concerned. Nevertheless, the relative position of the dual atoms clouds could be important for particular applications in the dual atom GMOT. Thus, the fabrication uniformity is more crucial than the conventional cooling atoms grating chip. In conclusion, this paper provides a valuable approach for designing multicolor GMOTs and offers guidance for future experiments involving multispecies atom cooling and trapping in GMOT systems. The successful construction of the dual-atom GMOT system enlarges the spectral line richness of compact cold atom systems, which is expected to improve the precision measurement accuracy and range. It can be used in cold atom clocks, quantum interferometers, and quantum electrometry. The combination with the metasurface technology can also be used to develop integrated compact quantum devices.

**Funding.** National Natural Science Foundation of China (12273087); Natural Science Foundation of Hubei Province (2024AFA071); Strategic Priority Research Program of the Chinese Academy of Sciences (XDB1070103); National Key Research and Development Program of China (2021YFF0603701).

**Acknowledgment.** Thanks to Weiming Lv for the grating chip fabrication and Yani Zuo for laser cooling technique discussion.

**Disclosures.** The authors declare no conflicts of interest.

**Data availability.** Data underlying the results presented in this paper are not publicly available at this time but may be obtained from the authors upon reasonable request.